\documentclass[12pt]{article}
\usepackage{amssymb}
\usepackage{amsmath}
\usepackage{color}
\usepackage{cancel}
\usepackage[affil-it]{authblk}
 
\usepackage{graphicx}

\newtheorem{theorem}{Theorem}
\newtheorem{prop}[theorem]{Proposition}

\newtheorem{cor}[theorem]{Corollary}

\newtheorem{remark}[theorem]{Remark}
\newenvironment{rem}{\begin{remark} \rm}{\end{remark}}
\newtheorem{example}[theorem]{Example}

\newenvironment{exa}{\begin{example} \rm}{\end{example}}

\def\d{{\mathrm d}}

\usepackage{cancel}

\newcommand{\RR}{{{\mathbb{R}}}}

\def\M{{\mathcal M}}
\def\Tr{\operatorname{Tr}}

\begin{document}
\title{Poisson quasi-Nijenhuis deformations\\of the canonical PN structure}
\date{} 

\author{G.\ Falqui${}^{1,4}$, I.\ Mencattini${}^2$, 
M.\ Pedroni${}^{3,4}$}

\affil{
{\small  $^1$Dipartimento di Matematica e Applicazioni, Universit\`a di Milano-Bicocca, Italy
}\\
{\small  gregorio.falqui@unimib.it 
}\\
\medskip
{\small $^2$Instituto de Ci\^encias Matem\'aticas e de Computa\c c\~ao,  Universidade de S\~ao Paulo, Brazil}\\
{\small igorre@icmc.usp.br 
}\\
\medskip
{\small $^3$Dipartimento di Ingegneria Gestionale, dell'Informazione e della Produzione,  Universit\`a di Bergamo, Italy}\\
{\small marco.pedroni@unibg.it 
}\\
\medskip
{\small  $^4$INFN, Sezione di Milano-Bicocca, Piazza della Scienza 3, 20126 Milano, Italy}
}

\maketitle
\abstract{\noindent
We present a result which allows us to deform a Poisson-Nijenhuis manifold into a Poisson quasi-Nijenhuis manifold by means of a closed
2-form. Under an additional assumption, the deformed structure is also Poisson-Nijenhuis. We apply this result to show that the canonical 
Poisson-Nijenhuis structure on 
${\mathbb R}^{2n}$ gives rise to both the Poisson-Nijenhuis structure of the open (or non periodic) $n$-particle Toda lattice, introduced by 
Das and Okubo \cite{DO}, 
and the Poisson quasi-Nijenhuis structure of the closed (or periodic) $n$-particle Toda lattice, described in our recent work \cite{FMOP2020}.}


\baselineskip=0,6cm

\section{Introduction}
The notion of Poisson-Nijenhuis (PN) manifold \cite{MagriMorosiRagnisco85,KM} was introduced in connection with the theory of integrable systems. Such a manifold is endowed with a Poisson tensor $\pi$ and with a tensor field $N$ of type $(1,1)$, sometimes called ``recursion operator",
which is torsionless and compatible (see Section \ref{sec:PN-PqN}) with 
$\pi$. It is a bi-Hamiltonian manifold, and the traces $H_k$ of the powers 
of $N$ 
are in involution with respect to both Poisson brackets induced by the Poisson tensors. 
A Poisson-Nijenhuis structure for the open (or non periodic) $n$-particle Toda lattice (see \cite{Perelomov-book} and references therein) was presented in \cite{DO} (see also \cite{Damianou04,MorosiPizzocchero96,Okubo90}). 

A generalization of PN manifolds was introduced in \cite{SX}, where a Poisson quasi-Nijenhuis (PqN) manifold was defined to be a Poisson manifold with a compatible tensor field $\widehat N$ of type $(1,1)$, whose torsion need not vanish but is controlled by a suitable 3-form
$\phi$. In \cite{FMOP2020} we found an application of this geometrical object to finite-dimensional integrable systems. More precisely, we 
obtained a rather stringent sufficient condition 
to transform, 
by means of a 2-form $\Omega$, a 
PN manifold $(\M,\pi,N)$ into a PqN manifold $(\M,\pi,\widehat N,\phi)$. We also found suitable compatibility conditions between $\pi$, $N$ and $\Omega$ entailing that the functions $H_k=\frac1{2k}\Tr({\widehat N}^k)$ are in involution.
Then we interpreted the well known integrability of the closed Toda lattice in this framework, showing that its integrals of motion are the traces of the powers of a suitable tensor field $\widehat N$ of type $(1,1)$, which is a deformation of the Das-Okubo recursion operator $N$ of the open Toda lattice. 

In this paper we give a stronger result than the above mentioned deformation scheme, in the sense that the only condition on the 2-form $\Omega$, to obtain a PqN manifold, is $\d\Omega=0$. Under an additional assumption, the deformed structure is a PN manifold. Then we apply this result to show that the ``canonical" PN structure on ${\mathbb R}^{2n}$ (corresponding to the free-particle case) gives rise to both the PN structure \cite{DO} 
of the open 
Toda lattice and the PqN structure \cite{FMOP2020} of the closed 
Toda lattice.

The organization of this paper is the following. In Section \ref{sec:PN-PqN} we recall the definitions of PN and PqN manifold. Section \ref{igor-connection} is devoted to the (improved version of the) above mentioned deformation theorem, with further considerations on the case where $\pi$ is non degenerate. This result is applied in Section \ref{defo-can} to a wide class of mechanical systems, including the open and closed 
$n$-particle Toda systems, whose PN and PqN structures are interpreted as deformations of the canonical PN structure on ${\mathbb R}^{2n}$.


\par\smallskip\noindent
{\bf Acknowledgments.}
We thank Giovanni Ortenzi for useful discussions. MP thanks the Department of Mathematics and its Applications of the University of Milano-Bicocca for its hospitality. 
This project has received funding from the European Union's Horizon 2020 research and innovation programme under the Marie Sk{\l}odowska-Curie grant no 778010 {\em IPaDEGAN}. All authors gratefully acknowledge the auspices of the GNFM Section of INdAM under which part of this work was carried out.


\section{Poisson quasi-Nijenhuis manifolds}
\label{sec:PN-PqN}
Let $N:T\M\to T\M$ be a $(1,1)$ tensor field on a manifold $\M$.
It is well known that its {\it Nijenhuis torsion\/} is defined as 
\begin{equation}
\label{tndef1}
T_N(X,Y)=[NX,NY]-N\left([NX,Y]+[X,NY]-N[X,Y]\right).
\end{equation}
We also recall that, given a $p$-form $\alpha$, with $p\ge 1$, one can construct another $p$-form $i_N\alpha$ as \begin{equation}
\label{iNalpha}
i_N\alpha(X_1,\dots,X_p)=\sum_{i=1}^p \alpha(X_1,\dots,NX_i,\dots,X_p),
\end{equation}
and that $i_N$ is a derivation 
of degree zero (if $i_N f=0$ for all functions $f$). 

If $\pi$ is a Poisson bivector on $\M$ and $\pi^\sharp:T^*\M\to T\M$ is defined by $\langle \beta,\pi^\sharp\alpha\rangle=\pi(\alpha,\beta)$, then 
$\pi$ and $N$ are said to be {\it compatible\/} \cite{MagriMorosiRagnisco85} if 
\begin{equation}
\label{N-P-compatible}
\begin{split}
&N\pi^\sharp=\pi^\sharp N^*\,,\qquad
\mbox{where $N^*:T^*\M\to T^*\M$ is the transpose of $N$;}\\
&L_{\pi^\sharp\alpha}(N) X-\pi^\sharp
L_{X}(N^*\alpha)+\pi^\sharp
L_{NX}\alpha=0,\qquad\mbox{for all 1-forms $\alpha$ and vector fields $X$.}
\end{split}
\end{equation}
In \cite{SX} a {\it Poisson quasi-Nijenhuis (PqN) manifold\/} was defined as a quadruple $(\M,\pi,N,\phi)$ such that:
\begin{itemize}
\item the Poisson bivector $\pi$ and the $(1,1)$ tensor field $N$  
are compatible;
\item the 3-forms $\phi$ and $i_N\phi$ are closed;
\item $T_N(X,Y)=\pi^\sharp\left(i_{X\wedge Y}\phi\right)$ for all vector fields $X$ and $Y$, where $i_{X\wedge Y}\phi$ is the 1-form defined as $\langle i_{X\wedge Y}\phi,Z\rangle=\phi(X,Y,Z)$.
\end{itemize}
A slightly more general definition of PqN manifold was recently proposed in \cite{BursztynDrummond2019}.

If $\phi=0$, then the torsion of $N$ vanishes and $\M$ becomes a {\it Poisson-Nijenhuis manifold} (see \cite{KM} and references therein). In this case, the bivector field $\pi_N$ defined by $\pi_N^\sharp=N\pi^\sharp$ is a Poisson tensor compatible with $\pi$, so that $\M$ is a bi-Hamiltonian manifold. 
Moreover, the functions
\begin{equation}
\label{tracce}
H_k=\frac1{2k}\Tr(N^k),\qquad k=1,2,\dots,
\end{equation}
satisfy $\d H_{k+1}=N^* \d H_{k}$, entailing
their involutivity with respect to both Poisson brackets induced by $\pi$ and $\pi_N$. 

In \cite{FMOP2020} we called {\em involutive\/} a PqN manifold such that the traces (\ref{tracce}) of the powers of $N$ are in involution (with respect to the unique Poisson bracket defined on $\M$, i.e., the one associated with $\pi$) and we observed that there are non involutive PqN manifolds. Moreover, we found some hypotheses to be added to obtain a class of involutive PqN manifolds, and we gave an application to the closed Toda lattice, whose PqN structure has been defined as a suitable deformation (see next section) of the PN structure \cite{DO} of the open Toda lattice. 


\section{Deformations of PN manifolds}
\label{igor-connection}

In this section we first present a few basic 
facts about the theory of PN and PqN manifolds from the view-point of 
differential graded Lie algebras. Then we prove a result which gives a sufficient condition to deform a PN structure into a PqN one.  

First of all, we recall that, given a tensor field $N:T\M\to T\M$, the usual Cartan differential can be modified as follows,
\begin{equation}
\label{eq:dN}
\begin{split}
(\d_N\alpha)(X_0,\dots,X_q)&=
\sum_{j=0}^q(-1)^j L_{NX_j}\left(\alpha(X_0,\dots,\widehat{X}_j,\dots,X_q)\right)\\&
+\sum_{i<j}(-1)^{i+j}\alpha([X_i,X_j]_N,X_0,\dots,\widehat{X}_i,\dots,\widehat{X}_j,\dots,X_q),
\end{split}
\end{equation}
where $\alpha$ is a $q$-form, the $X_i$ are vector fields,
and 
$[X,Y]_N=[NX,Y]+[X,NY]-N[X,Y]$. Note that 
$\d_Nf=N^* \d f$ for all $f\in C^\infty(\M)$. Moreover, 
\begin{equation}
\label{eq:dNd}
\d_N=i_N\circ \d-\d\circ i_N,
\end{equation}
where $i_N$ is given by (\ref{iNalpha}), and consequently $\d\circ \d_N+\d_N\circ \d=0$. 
Finally, $
\d_N
^2 =0$ if and only if the torsion of $N$ vanishes.

We also remind that one can define a Lie bracket between 1-forms on a Poisson manifold $(\M,\pi)$ as
\begin{equation}
\label{eq:liealgpi}
[\alpha,\beta]_\pi=L_{\pi^\sharp\alpha}\beta-L_{\pi^\sharp\beta}\alpha-\d\langle\beta,\pi^\sharp\alpha\rangle,
\end{equation}
and that this Lie bracket 
can be uniquely extended to all forms on $\M$ in such a way that, if $\eta$ is a $q$-form and $\eta'$ is a $q'$-form, then 
$[\eta,\eta']_\pi$ is a $(q+q'-1)$-form and 
\begin{itemize}
\item[(K1)] $[\eta,\eta']_\pi=-(-1)^{(q-1)(q'-1)}[\eta',\eta]_\pi$; 
\item[(K2)] $[\alpha,f]_\pi=i_{\pi^\sharp\alpha}\,\d f=\langle \d f,\pi^\sharp\alpha\rangle$ for all $f\in C^\infty(\M)$ and for all 1-forms $\alpha$;
\item[(K3)] 
$[\eta,\cdot]_\pi$ 
is a derivation of degree $q-1$ of the wedge product, that is, 
for any differential form $\eta''$,
\begin{equation}
\label{deriv-koszul}
[\eta,\eta'\wedge\eta'']_\pi=[\eta,\eta']_\pi\wedge\eta''+(-1)^{(q-1)q'}\eta'\wedge[\eta,\eta'']_\pi.
\end{equation}
\end{itemize}
This extension is a {\it graded\/} Lie bracket, in the sense that (besides (K1)) the graded Jacobi identity holds:
\begin{equation}
\label{graded-jacobi}
(-1)^{(q_1-1)(q_3-1)}[\eta_1,[\eta_2,\eta_3]_\pi]_\pi+(-1)^{(q_2-1)(q_1-1)}[\eta_2,[\eta_3,\eta_1]_\pi]_\pi+(-1)^{(q_3-1)(q_2-1)}[\eta_3,[\eta_1,\eta_2]_\pi]_\pi=0
\end{equation}
where $q_i$ is the degree of $\eta_i$.
It is sometimes called the Koszul bracket --- see, e.g., \cite{FiorenzaManetti2012} and references therein. We warn the reader that the Koszul bracket we used in \cite{FMOP2020} is the opposite of the one used here, since a minus sign in (K2) was inserted. For future reference, we remark that (K2) holds for any differential form $\eta$ and for all $f\in C^\infty(\M)$. More precisely,
\begin{equation}
\label{K2pertutte}
[f,\eta]_\pi=i_{\pi^\sharp \d f}\,\eta.
\end{equation}
Indeed, $[f,\cdot]_\pi$ and $i_{\pi^\sharp \d f}$ coincide on 0-forms (they both vanish) and on 1-forms (thanks to (K1) and (K2)), and they are derivations 
of degree $-1$ with respect to the wedge product.

It was proved in \cite{YKS96} that the compatibility conditions (\ref{N-P-compatible}) between a Poisson tensor $\pi$ and a tensor field $N:T\M\to T\M$ hold if and only if 
$\d_N$ is a derivation of $[\cdot,\cdot]_\pi$, 
that is,
\begin{equation}
\label{deriv-wedge}
\d_N[\eta,\eta']_\pi=[\d_N\eta,\eta']_\pi+(-1)^{(q-1)}[\eta,\d_N\eta']_\pi
\end{equation}
if $\eta$ is a $q$-form and $\eta'$ is any differential form. In particular, taking $N={\rm Id}$, one has that the Cartan differential $\d$ is always a derivation of  
$[\cdot,\cdot]_\pi$.
\begin{rem}
It is worth noting that (K2) is a consistency requirement. In fact, for $f,g\in C^\infty(\mathcal M)$ one has that
\begin{align}
\d[\d g,f]_\pi\stackrel{\eqref{deriv-wedge}}{=}[\d g,\d f]_\pi\stackrel{\eqref{eq:liealgpi}}{=}\d\{g,f\}.  
\end{align}
On the other hand,
\begin{align}
\d[\d g,f]_\pi\stackrel{\rm (K2)}{=}\d\left(i_{\pi^\sharp
\d g}(\d f)\right)=\d\{g,f\}.
\end{align}
\end{rem}

The following result generalizes Theorem 3 in \cite{FMOP2020}.
\begin{theorem}
\label{thm:gim}
\label{thm:def-conv}
Let $(\M,\pi,N)$ be a PN manifold and let $\Omega$ be a closed 2-form. 
If $\widehat N=N+\pi^\sharp\,\Omega^\flat$, where $\Omega^\flat:T\M\to T^*\M$ is defined as usual by $\Omega^\flat(X)=i_X\Omega$, and 
\begin{equation}
\label{eq:condiphi}
\phi=\d_N\Omega+\frac{1}{2}[\Omega,\Omega]_\pi,
\end{equation} 
then $(\M,\pi,\widehat N,\phi)$ is a PqN manifold. In particular, if 
\begin{equation}\label{eq:homeq}
\d_N\Omega+\frac{1}{2}[\Omega,\Omega]_\pi=0,
\end{equation}
then $(\M,\pi,\widehat N)$ is a PN manifold.
\end{theorem}
{\bf Proof.} First we prove the compatibility between $\pi$ and $\widehat N$. If $\d_{\widehat N}=\d_N+\d_{\pi^\sharp\,\Omega^\flat}$ is the differential defined by $\widehat N=N+\pi^\sharp\,\Omega^\flat$, 
we observe that 
\begin{equation}
\d_{\pi^\sharp\,\Omega^{\flat}}=[\Omega,\cdot]_\pi.\label{eq:monella}
\end{equation}
Indeed, both the left and the right-hand side of the previous formula 
are graded derivations 
with respect to the wedge product, anti-commuting with $\d$ (since $\d\Omega=0$), and coinciding on 0-forms. To prove the last assertion, we first notice that, for all $f\in C^\infty(\mathcal M)$,
\[
\d_{\pi^\sharp\,\Omega^\flat}f=(\pi^\sharp\,\Omega^\flat)^\ast(\d f)=\Omega^\flat(\pi^\sharp\,\d f)=i_{\pi^\sharp\,\d f}\,\Omega=[\Omega,f]_{\pi},
\]
where we used (\ref{K2pertutte}) and (K1) in the last equality. Hence, since $\d_N$ and $[\Omega,\cdot]_\pi$ are both derivations of $[\cdot,\cdot]_\pi$, it follows at once that $\d_{\widehat N}$ is also a derivation of $[\cdot,\cdot]_\pi$, yielding the compatibility between $\widehat N$ and $\pi$.
Moreover,  
\begin{equation}
\d_{\widehat N}\phi=\d_N\phi+[\Omega,\phi]_\pi\stackrel{\eqref{eq:condiphi}}{=}\d_N\left(\d_N\Omega+\frac{1}{2}[\Omega,\Omega]_\pi\right)+
\left[\Omega,\d_N\Omega+\frac{1}{2}[\Omega,\Omega]_\pi\right]_\pi=\frac{1}{2}[\Omega,[\Omega,\Omega]_\pi]_\pi=0
\end{equation} 
thanks to $\d_N^2=0$, the Leibniz rule (\ref{deriv-wedge}) for $\d_N$, the commutation rule (K1), and the graded Jacobi identity (\ref{graded-jacobi}). Then it follows from (\ref{eq:dNd}) and $\d\phi=0$ that $i_{\widehat N}\phi$ is closed.
Finally, with the same argument used to show (\ref{eq:monella}), we can prove that $\d_{\widehat N}^2=[\phi,\cdot]_\pi$. Indeed, using again 
$\d_N^2=0$, we obtain
\begin{equation}
\label{d_N^2}
\begin{aligned}
\d_{\widehat N}^2f&=(\d_N+[\Omega,\cdot]_\pi)^2 f= (\d_N+[\Omega,\cdot]_\pi)(\d_Nf+[\Omega,f]_\pi)\\
&=\d_N^2f+\d_N[\Omega,f]_\pi+[\Omega,\d_Nf]+[\Omega,[\Omega,f]_\pi]_\pi\\
&\stackrel{\eqref{deriv-wedge}}{=}[\d_N\Omega,f]_\pi-\cancel{[\Omega,\d_Nf]_\pi}+\cancel{[\Omega,\d_Nf]_\pi}+[\Omega,[\Omega,f]_\pi]_\pi\\
&\stackrel{\eqref{graded-jacobi}}{=}[\d_N\Omega,f]_\pi+
\frac{1}{2}[[\Omega,\Omega]_\pi,f]_\pi
=[\phi,f]_\pi.
\end{aligned}
\end{equation}
To conclude the proof of the first assertion in the theorem, it suffices to use the fact (see \cite{SX}) that, for any 3-form $\phi$,
\begin{equation}
\label{iff:SX}
\d_{\widehat N}^2=[\phi,\cdot]_\pi\quad\mbox{if and only if}\quad 
\left\{\begin{array}{l}
T_{\widehat N}(X,Y)=\pi^\sharp\left(i_{X\wedge Y}\phi\right)\quad\mbox{ for all vector fields $X,Y$}\\
\noalign{\medskip}
i_{(\pi^\sharp\alpha)\wedge(\pi^\sharp\beta)\wedge(\pi^\sharp\gamma)}(\d\phi)=0\quad\mbox{ for all 1-forms $\alpha,\beta,\gamma$}
\end{array}\right.
\end{equation}
The proof of the second assertion simply follows, recalling that a PqN manifold $(\mathcal M,\pi,\widehat N,\phi)$ whose 3-form $\phi$ vanishes is a PN manifold.
\hfill$\square$\medskip

We remark that in a similar way one can prove the following result (see \cite{FMOP2020}):
If $(\mathcal M,\pi,N,\phi)$ is a PqN manifold, $\Omega
$ is a closed 2-form such that 
\[
-\phi=\d_N\Omega+\frac{1}{2}[\Omega,\Omega]_\pi,    
\]
and $\widehat N=N+\pi^\sharp\,\Omega^\flat$, then $(\mathcal M,\pi,\widehat N)$ is a PN manifold. 

We also notice that, in the case where $N={\rm Id}$, equation \eqref{eq:homeq} was 
studied in \cite{LWX}, 
in the framework of the theory of Manin triples for Lie algebroids. 
Starting from a Poisson manifold $(\mathcal M,\pi)$, 
it was shown that every solution of 
\begin{equation}\label{eq:homeq-id}
\d\Omega+\frac{1}{2}[\Omega,\Omega]_\pi=0
\end{equation}
defines a Dirac subbundle $\Gamma_\Omega\subset T^\ast\mathcal M\oplus T\mathcal M$ transversal to $T^\ast\mathcal M$, and that
every solution of
\begin{equation}
\d\Omega=0\quad\text{and}\quad[\Omega,\Omega]_\pi=0\label{complementary}
\end{equation}
defines 
a PN structure on $\mathcal M$. As we will see in 
Corollary \ref{cor:inv}, 
the latter result becomes an equivalence under the further hypothesis that $\pi$ is non degenerate.

\subsection{Deformations of symplectic-Nijenhuis manifolds}

In this subsection we consider the important, though particular, case of a PN manifold $(\mathcal M,\pi,N)$ whose Poisson tensor is non degenerate, 
i.e., such that $\pi^\sharp
$ is invertible, so that it defines a symplectic form $\omega$ via the 
identity
\begin{equation}
\label{eq:convomega}
\pi^\sharp\omega^\flat=-{\rm Id}.
\end{equation}
The relevance of this case stems from the theory of classical integrable systems, where $\mathcal M$ is a cotangent bundle endowed with its canonical symplectic structure.
\begin{rem}
The defining relation \eqref{eq:convomega} is the same chosen in \cite{Complementary}, but we alert the reader that it is not universally adopted. For example, the authors of \cite{SX} chose $\pi^\sharp\omega^\flat=+{\rm Id}$.
\end{rem}

Following \cite{SX,Complementary}, we introduce the notion of a \emph{symplectic-Nijhenuis manifold}, which is a Poisson-Nijenhuis manifold $(\mathcal M,\pi,N)$ whose Poisson tensor $\pi$ is non degenerate. Hereafter a symplectic-Nijenhuis manifold will be denoted by a triple $(\mathcal M,\omega,N)$ where $\omega$ is the symplectic form defined in \eqref{eq:convomega}. In this more specialized framework, the properties of the PN structure find a covariant analogue. In particular, 
the bilinear form $\omega_N:T\mathcal M\times T\mathcal M\rightarrow\mathbb R$ defined by 
$$
\omega_N(X,Y)=\omega(N X,Y),
$$
where $X,Y$ are vector fields on $\mathcal M$, is skew-symmetric; hence it defines a 2-form on $M$, called the \emph{associated 2-form} of $(\mathcal M,\omega,N)$. 
The following result provides a characterization of the associated 2-forms, see \cite[Theorem 2.1]{Complementary}. 

\begin{prop}\label{pro:vai}
The associated 2-form of a symplectic-Nijenhuis manifold $(\mathcal M,\omega,N)$ satisfies the 
conditions
\begin{equation}
\d\omega_N=0 \quad\text{and}\quad  [\omega_N,\omega_N]_\pi=0,\label{eq:vais}
\end{equation}
where $\pi$ is the Poisson tensor corresponding to $\omega$.
On the other hand, if a 2-form $\omega_N$ on a symplectic manifold $(\mathcal M,\omega)$ satisfies \eqref{eq:vais}, 
then $\omega_N$ is the associated 2-form of the symplectic-Nijenhuis manifold $(M,\omega,N)$, where 
$N=\left(\omega^\flat\right)^{-1}\omega_N^\flat=-\pi^\sharp\omega_N^\flat$. 
\end{prop}

\begin{rem}
\label{rem:kozul2form}
For future reference, we notice that (\ref{eq:monella}) entails
\begin{equation}
[\omega
,\cdot]_\pi=\d_{\pi^\sharp\omega
^\flat}=\d_{-{\rm Id}}=-\d,\qquad
[\omega_N,\cdot]_\pi=\d_{\pi^\sharp\omega_N^\flat}=-\d_{N}.
\end{equation}
\end{rem}

Now we show that the sufficient condition contained in the PN part of Theorem \ref{thm:gim} is also necessary when $\pi$ is non degenerate.
\begin{prop}
\label{prop:con}
Let $(\mathcal M,\omega,N)$ be a symplectic-Nijenhuis manifold, $\widehat N$ a (1,1) tensor field, and $\Omega$ the (0,2) tensor field defined by 
\begin{equation}
\label{eq:N}
\widehat N=N+\pi^\sharp\Omega^\flat,
\end{equation}
where $\pi$ is given by (\ref{eq:convomega}). Then $\widehat N$ is torsionless and compatible with $\pi$ if and only if 
$\Omega$ is a closed 2-form such that 
\begin{equation}
\d_{N}\Omega+\frac{1}{2}[\Omega,\Omega]_\pi=0.\label{eq:mc}
\end{equation}
\end{prop}
{\it Proof.} If the above mentioned conditions on $\Omega$ hold, then Theorem \ref{thm:gim} implies 
that $(\mathcal M,\pi,\widehat N)$ is a PN manifold.\\
Viceversa, multiplying \eqref{eq:N} on the left by the inverse of $\pi^\sharp$ and considering the associated 2-forms $\omega_N$ and 
$\omega_{\widehat N}$, one obtains
\begin{equation}
\label{omegadiff}
\Omega=\omega_{N}-\omega_{\widehat N},
\end{equation}
showing that $\Omega$ is a closed 2-form thanks to Proposition \ref{pro:vai}. Moreover, from (\ref{d_N^2}) and $d_N^2=d_{\widehat N}^2=0$ 
we have that $\phi=\d_N\Omega+\frac{1}{2}[\Omega,\Omega]_\pi$ satisfies 
$[\phi,f]_\pi=0$ for all functions $f$. Then (\ref{K2pertutte}) implies that $i_{\pi^\sharp\d f}\phi=0$, so that $\phi=0$ follows from the fact that $\pi$ is non degenerate.
\hfill$\square$\medskip

Putting $N={\rm Id}$, we obtain the following corollary, to be compared with the results of \cite{LWX}, Section 6.

\begin{cor}\label{cor:inv}
Let $(\mathcal M,\omega)$ be a symplectic manifold and let $\pi$ be the Poisson tensor defined by $\omega$, see 
\eqref{eq:convomega}. A tensor field $\widehat N$ of type (1,1) is torsionless and compatible with $\pi$ if and only if the $(0,2)$ tensor field
$\Omega$ defined by 
\begin{equation}
\widehat N={\rm Id}+\pi^\sharp\Omega^\flat\label{eq:ag}
\end{equation}
is a 2-form satisfying the conditions
\begin{equation}
\label{eq:ag1}
\d\Omega=0\quad\text{and}\quad [\Omega,\Omega]_\pi=0.
\end{equation}
\end{cor}

In the next section we will deform the so called {\it canonical PN structure}, i.e., the symplectic-Nijenhuis manifold
$(\mathbb R^{2n},\pi,N
)$, where $\pi=\sum_{i=1}^n \partial_{p_i}\wedge\partial_{q_i}$ is the canonical Poisson tensor and 
\begin{equation}
N
=\sum_{i=1}^np_i(\partial_{q_i}\otimes \d q_i+\partial_{p_i}\otimes \d p_i)\label{eq:canten}
\end{equation}
is easily seen to be torsionless and compatible with $\pi$. It was shown in \cite{Turiel} that this is the local form of any 
$2n$-dimensional symplectic-Nijenhuis manifold, provided that the eigenvalues of the recursion operator are $n$ independent functions $p_1,\dots,p_n$. In this context, 
$(q_1,\dots,q_n,p_1,\dots,p_n)$ were called Darboux-Nijenhuis coordinates and used as separation variables in \cite{FP03}.
In the following example we show that $N
$ itself can be seen as a deformation. 

\begin{example}
\label{exa:defo-ide}
{\rm The above mentioned symplectic-Nijenhuis manifold $(\mathbb R^{2n},\pi,N
)$ can be obtained, using Corollary \ref{cor:inv}, as a deformation of the symplectic manifold $(\mathbb R^{2n}, \omega)$, where  
$\omega
=\sum_{i=1}^n \d p_i\wedge \d q_i$ is the canonical symplectic form, i.e., the one corresponding to $\pi$.
Indeed, let us consider the 2-form 
\begin{equation}
\label{Omega-c}
\Omega
=\sum_{i=1}^n\d p_i\wedge \d q_i-\sum_{i=1}^np_i\d p_i\wedge \d q_i.
\end{equation}
Then $N
={\rm Id}+\pi^\sharp\,\Omega
^\flat$, and 
$\Omega
$ 
can be checked to be a solution of \eqref{eq:ag1}. 
Moreover, observe that, according to (\ref{omegadiff}), $\Omega
=\omega
-\omega_N$, where $\omega_N=\sum_{i=1}^np_i\d p_i\wedge \d q_i$ is the 2-form 
associated to $N$.
}
\end{example}

The final part of this section is devoted to two remarks on Corollary \ref{cor:inv}. 
\begin{rem}
If $(\mathcal M,\omega,N)$ is a symplectic-Nijenhuis manifold, then the 2-form $\Omega$ defined by $N=\text{Id}+\pi^\sharp\Omega^\flat$
satisfies (\ref{eq:ag1}). One can show that the pair $(\pi,\Omega)$ endows $\mathcal M$ with a 
$P\Omega$ structure 
(see \cite{MagriMorosiRagnisco85,YKS-Rubtsov}), in the sense that $\pi$ is a Poisson tensor, $\Omega$ is closed, and 
$\Omega^\flat\pi^\sharp\Omega^\flat$ is closed too. Indeed, $(\mathcal M,\omega,\pi^\sharp\Omega^\flat)$ is also a symplectic-Nijenhuis manifold, so that 
the same is true for $(\mathcal M,\omega,(\pi^\sharp\Omega^\flat)^2)$. But the 2-form associated to this latter is $-\Omega^\flat\pi^\sharp\Omega^\flat$, and this is closed by Proposition \ref{pro:vai}.\\
Since $\Omega^\flat\pi^\sharp\Omega^\flat$ is closed, one can easily check that the pair $(\Omega,N)$ endows $\mathcal M$ with an 
$\Omega N$ structure 
\cite{Magri-Morosi}, in the sense that $\Omega$ and $\Omega^\flat N$ are both closed, and $\Omega^\flat N=N^*\Omega^\flat$. It follows that 
$(\Omega_{N^k},N)$ is an $\Omega N$ structure too, where $\left(\Omega_{N^k}\right)^\flat=\Omega^\flat N^k$. In particular, all the 2-forms 
$\Omega_{N^k}$ are closed.\\
Considering the symplectic-Nijenhuis manifold $(\mathcal M,\omega,N^k)$, one can introduce the closed 2-form $\Omega_k$ defined by 
$N^2={\rm Id}+\pi^\sharp\Omega_k^\flat$ and check that, for all $k\geq 1$,
\[
\Omega_k=\sum_{l=0}^{k-1}\Omega_{N^l}.
\]
Indeed, a simple induction shows that
\[
N^k=\text{Id}+\pi^\sharp\left(\sum_{l=0}^{k-1}\Omega^\flat {N}^l\right).
\]
\end{rem}

\begin{rem}
It is well known (see, e.g., \cite{Magri-Morosi,Bonechi}) that the closed 1-forms $\alpha$ on a symplectic-Nijenhuis manifold 
$(\mathcal M,\omega,N)$ that fulfills $\d_N\alpha=0$ (or, equivalently, $\d(N^*\alpha)=0$) play a fundamental role in the applications 
to integrable systems. Such forms coincide with the (closed) ones such that $L_{\pi^\sharp\alpha}\Omega=0$.  Indeed, 
$$
L_{\pi^\sharp\alpha}\Omega=\d
(i_{\pi^\sharp\alpha}\Omega
)=\d
(\Omega^\flat\pi^\sharp\alpha
)=\d
(N^*\alpha
).
$$
In the case 
discussed in and before Example \ref{exa:defo-ide}, it can be easily checked that the closed 1-form such that $L_{\pi^\sharp\alpha}\Omega=0$ are the differentials of the functions depending only on $p_1,\dots,p_n$. The algebra of these functions is abelian (with respect to the Poisson bracket) and is generated by the traces of the powers of $N$.
\end{rem}


\section{Deformations of the canonical PN structure}\label{defo-can}

In this section we start from the canonical PN structure $(\mathbb R^{2n},\pi,N)$, described before Example \ref{exa:defo-ide}, and we apply Theorem \ref{thm:def-conv} to recover the PN (respectively PqN) structure of the open (respectively closed) Toda lattice. The traces of the powers of $N$ are simply $H_k=\frac1{2k}\Tr(N^k)=\sum_{i=1}^n p_i^k$, so that from the physical point of view the starting point is nothing but the free-particle case. On the contrary, in \cite{FMOP2020} we used a preliminary version of Theorem \ref{thm:def-conv} to construct the PqN structure of the closed Toda lattice out of  the PN structure of the open one.

We consider the 2-form 
\begin{equation}
\label{Omega-Vij}
\Omega=\sum_{i<j}\left(V_{ij}(q_i-q_j)\d q_j\wedge \d q_i+\d p_j\wedge \d p_i\right),
\end{equation}
where $V_{ij}$ are $C^\infty$ functions of a single variable. One has that $\Omega=\d\theta$, where
\begin{equation}
\theta=\sum_{i<j}\left(-{\tilde V}_{ij}\d q_i+p_j \d p_i\right)
\end{equation}
and ${\tilde V}_{ij}$ is a primitive function of ${V}_{ij}$. The deformed tensor field $\widehat N=N+\pi^\sharp\,\Omega^\flat$ turns out to be
\begin{equation}
\label{N-hat_Vij}
\widehat N=\sum_{i=1}^{n}p_i\left(\partial_{q_i}\otimes \d q_i +\partial_{p_i}\otimes \d p_i\right)
+\sum_{i<j}\left(\partial_{q_i}\otimes \d p_j-\partial_{q_j}\otimes \d p_i\right)
+\sum_{i<j}V_{ij}\left(\partial_{p_{j}}\otimes \d q_i-\partial_{p_i}\otimes \d q_{j}\right).
\end{equation}
We refer to Subsection \ref{subsec-2particle} for the matrix form of $\widehat N$ in the case $n=2$, from which the general expression is easily guessed.
Theorem \ref{thm:def-conv} entails that $(\RR^{2n},\pi,\widehat N,\phi)$ is a PqN manifold, where 
$\phi=\d_N\Omega+\frac12[\Omega,\Omega]_\pi$. One can show that 
\begin{equation}
\label{phi_Vij}
\begin{aligned}
\d_N\Omega&=\sum_{i<j}V_{ij}\d q_i\wedge \d q_{j}\wedge\left(\d p_i+\d p_{j}\right),\\
[\Omega,\Omega]_\pi&=2\sum_{i<j}V'_{ij}\d q_i\wedge \d q_{j}\wedge
\sum_{k<l}\left((\delta_{il}-\delta_{jl})\d p_k+(\delta_{jk}-\delta_{ik})\d p_{l}\right),
\end{aligned}
\end{equation}
where $\delta_{ij}$ is the Kronecker delta.
Notice that $H_2=\frac14\Tr({\widehat N}^2)=\frac12\sum_{i=1}^{n}p_i^2+\sum_{i<j}V_{ij}(q_i-q_{j})$ is the Hamiltonian of $n$ interacting particles (with unit mass) and that $\{H_1,H_k\}=0$ for all $k$, since $\widehat N$ depends only on the differences $q_i-q_{j}$. However, in general $\{H_j,H_k\}\ne 0$, i.e., the PqN structure is not involutive (see \cite{FMOP2020} and the Calogero example below). 

\begin{example}
{\rm We consider the case where $V_{i,i+1}(x)=f_i\, {\rm e}^x  
$ for all $i=1,\dots,n-1$ and $V_{1n}(x)=f_n\, {\rm e}^{-x} 
$ (where the $f_i$ are some constants), the other $V_{ij}$ vanishing. The deformed tensor field $\widehat N
$ is given by
\begin{equation}
\label{N-closedToda}
\begin{split}
\widehat N&=\sum_{i=1}^{n}p_i\left(\partial_{q_i}\otimes \d q_i+\partial_{p_i}\otimes \d p_i\right)+
\sum_{i<j}\left(\partial_{q_i}\otimes \d p_j-\partial_{q_j}\otimes \d p_i\right)\\
&+\sum_{i=1}^{n-1}f_i\,{\rm e}^{q_i-q_{i+1}}\left(\partial_{p_{i+1}}\otimes \d q_i-\partial_{p_i}\otimes \d q_{i+1}\right)
-f_n\,{\rm e}^{q_n-q_{1}}\left(\partial_{p_{1}}\otimes \d q_n-\partial_{p_n}\otimes \d q_{1}\right),
\end{split}
\end{equation}
while 
$\phi=
2f_n\,\mathrm{e}^{q_n-q_1}\d q_1\wedge \d q_n\wedge\sum_{i=1}^n\d p_i
$. \\
If $f_i=1$ for all $i=1,\dots,n$, then we obtain the PqN structure of the closed Toda lattice, see \cite{FMOP2020}. 
The $H_k$ 
can be shown to be in involution since the PqN manifold is involutive. For example,
\begin{equation}
\label{traces-closedToda}
H_1=\frac12\Tr(\widehat N)=\sum_{i=1}^{n}p_i,\qquad H_2=\frac14\Tr({\widehat N}^2)=\frac12\sum_{i=1}^{n}p_i^2+\sum_{i=1}^{n-1}
\, {\rm e}^{q_i-q_{i+1}}+
\, {\rm e}^{q_n-q_{1}}
\end{equation}
are respectively the total momentum and the energy. \\
If 
$f_n=0$, then
\begin{equation}
\label{phi-openToda}
\phi=\d_N\Omega+\frac{1}{2}[\Omega,\Omega]_\pi=0,    
\end{equation}
and we obtain the PN structure introduced in \cite{DO} to study the open Toda lattice 
(with generic values of the constants $f_i$, for $i=1,\dots,n-1)$.}
\end{example}

\begin{rem}
Using the notations of Example \ref{exa:defo-ide}, we note that one can recover the Das-Okubo $(1,1)$ tensor field 
\begin{equation}
\label{N-openToda}
\begin{split}
\widehat N&=\sum_{i=1}^{n}p_i\left(\partial_{q_i}\otimes \d q_i+\partial_{p_i}\otimes \d p_i\right)+
\sum_{i<j}\left(\partial_{q_i}\otimes \d p_j-\partial_{q_j}\otimes \d p_i\right)\\
&+\sum_{i=1}^{n-1}f_i\,{\rm e}^{q_i-q_{i+1}}\left(\partial_{p_{i+1}}\otimes \d q_i-\partial_{p_i}\otimes \d q_{i+1}\right),
\end{split}
\end{equation}
which is simply (\ref{N-closedToda}) with $f_n=0$, by deforming the identity with the 2-form 
$\widehat{\Omega}=\Omega
+\Omega_1$,
where $\Omega
$ is given by (\ref{Omega-c}) and 
\begin{equation}
\label{Omega-DO}
\Omega_1=\sum_{i=1}^{n-1}{\rm e}^{q_i-q_{i+1}}\d q_{i+1}\wedge \d q_i+\sum_{i<j}\d p_j\wedge \d p_i,
\end{equation}
see (\ref{Omega-Vij}). This means that $\widehat N
={\rm Id}+\pi^\sharp\,\widehat{\Omega}^\flat$ and 
\[
\d\widehat{\Omega}+\frac{1}{2}[\widehat{\Omega},\widehat{\Omega}]_\pi=0.
\]
Note that since $\widehat{\Omega}$ is closed, one should have $[\widehat{\Omega},\widehat{\Omega}]_\pi=0$. To double check the latter, 
we recall that $\Omega
=\omega
-\omega_N$ is a solution of (\ref{eq:ag1}), so that 
\begin{equation}
[\widehat{\Omega},\widehat{\Omega}]_\pi=\cancel{[\Omega
,\Omega
]_\pi}
+2[\Omega
,\Omega_1]_\pi+[\Omega_1,\Omega_1]_\pi 
=\cancel{2[\omega,\Omega_1]_\pi}-2[\omega_N,\Omega_1]_\pi+
[\Omega_1,\Omega_1]_\pi,
\end{equation}
where $[\omega
,\Omega_1]_\pi=0$ since $[\omega
,\cdot]_\pi=-\d$, providing the required identity thanks to $[\omega_N,\cdot]_{\pi}=-\d_{N}$ (see Remark \ref{rem:kozul2form}) and 
the fact that $\Omega_1$ solves
(\ref{phi-openToda}).
\end{rem}

\begin{exa}
Another interesting particular case of (\ref{Omega-Vij}) and (\ref{N-hat_Vij}) is $V_{ij}(x)=x^{-2}$ for all $i,j$. We have that
$$H_2=\frac14\Tr({\widehat N}^2)=\frac12\sum_{i=1}^{n}p_i^2+\sum_{i<j}(q_i-q_{i+1})^{-2}$$ is the Hamiltonian of the rational Calogero system. However, one can check that the corresponding PqN manifold is involutive for $n=3$ but not for $n=4$. In other words, the powers of the traces of $\widehat N$ give the integral of motions of the Calogero system only for $n=3$. The question whether the general rational Calogero system can be framed within the theory of PqN manifolds is still open.
\end{exa}

\subsection{The case \boldmath${n=2}$}
\label{subsec-2particle}

In this subsection we present some explicit formulas for the trivial case $n=2$. We first notice that, given a bivector $\pi$ and a coordinate system 
$(x_1,\dots, x_n)$ on a manifold $\M$, we have that $X=\pi^\sharp\alpha$ if and only if $X^i=\pi^{ij}\alpha_i$. Since we prefer to use column rather than row vectors, whenever we write $\pi^\sharp=A$ and $A$ is a matrix, we mean that the $(i,j)$ entry of $A$ is $\pi^{ji}$. Also, if $\Omega$ is a 2-form and $\Omega^\flat=A$, then $A_{ij}=\Omega_{ji}$. For the same reason, when $N$ is a (1,1) tensor field and we write $N=A$, 
we mean that the $(i,j)$ entry of $A$ is $N_j^i$. 

We start with the canonical PN structure on $\RR^{4}$, defined by 
\begin{equation} 
\label{PNcan-4}
\pi^\sharp=\left( \begin {array}{cc|cc} 
0&0&1&0\\ 
0&0&0&1\\ 
\hline
-1&0&0&0\\ 
0&-1&0&0\\ 
\end{array}\right),\qquad
N=\left( \begin {array}{cc|cc} 
p_{{1}}&0&0&0\\ 
0&p_{{2}}&0&0\\ 
\hline
0&0&p_{{1}}&0\\ 
0&0&0&p_{{2}}\\ 
\end{array}\right)\,.
\end{equation}
Then we consider the 2-form 
$\Omega=V(q_1,q_2)\d q_2\wedge \d q_1+\d p_2\wedge \d p_1$, 
where $V$ is any $C^\infty$ function of two variables. One has that $\Omega=\d \theta$, where
$\theta={\tilde V}\d q_1+p_2 \d p_1$ 
and ${\tilde V}$ is any function such that $\partial_{q_2}{\tilde V}=V$. The deformed tensor field turns out to be
$$
\widehat N=N+\pi^\sharp\,\Omega^\flat=\left( \begin {array}{cc|cc} 
p_{{1}}&0&0&1\\ 
0&p_{{2}}&-1&0\\ 
\hline
0&-V&p_{{1}}&0\\ 
V&0&0&p_{{2}}\\ 
\end{array}\right)\,.
$$ 
One can check that 
\begin{equation}
\label{phi_2part}
\d_N\Omega
=V\d q_1\wedge \d q_{2}\wedge\left(\d p_1+\d p_{2}\right),
\quad
[\Omega,\Omega]_\pi
=2 \d q_1\wedge \d q_{2}\wedge\left((\partial_{q_2}V) \d p_1-(\partial_{q_1}V) \d p_2\right),
\end{equation}
and apply Theorem \ref{thm:def-conv} to show that $(\RR^{4},\pi,\widehat N,\phi)$ is a PqN manifold, where 
$\phi=\d_N\Omega+\frac12[\Omega,\Omega]_\pi$. Notice that $\phi=0$ (i.e., the manifold is PN) if and only if $V=f_1{\rm e}^{q_1-q_2}$ for some constant $f_1$, corresponding to the 2-particle Toda system.

It is immediate to check that this PqN manifold is involutive if and only if $V(q_1,q_2)$ depends only on the difference $q_1-q_2$. 
However, one of the hypotheses of the involutivity Theorem 6 in \cite{FMOP2020} is that 
$[\Omega,\Omega]_\pi=0$, and one can see from the second of (\ref{phi_2part}) that this is true if and only if $V$ is constant.
Similar considerations hold for the other hypotheses.
Hence this theorem cannot be applied to the very simple example $V=V(q_1-q_2)$.
In particular, the well known integrability of the Toda lattice associated to the simple Lie algebra $B_2$, whose potential is 
$V(q_1,q_2)={\rm e}^{q_1-q_{2}}+{\rm e}^{q_2}$, cannot be interpreted with the help of the PqN manifold $(\RR^{4},\pi,\widehat N,\phi)$. We leave for future investigations how to suitably modify the 2-form $\Omega$, and we close our paper with an application of Proposition \ref{prop:con} to the case of the so called orthogonal Toda systems.

\begin{exa}
A bi-Hamiltonian formulation for the \emph{open orthogonal Toda systems} was described in \cite{NDC-D}, see also \cite{Damianou04rev,Damianou99}. In particular, the authors of the (first) above cited reference found 
a Poisson tensor $\widehat{\pi}$ which, together with the canonical one $\pi$, forms a PN structure for the Toda systems whose Hamiltonians are
\begin{equation}
H(p,q)=\frac{1}{2}\sum_{i=1}^np^2_i+\sum_{i=1}^{n-1}{\rm e}^{q_i-q_{i+1}}+{\rm e}^{kq_n},
\end{equation}
where $k=1$ (respectively, $k=2$) for the Lie algebras $B_n$ (respectively, $C_n$). The framework of \cite{NDC-D} is the same of 
Proposition \ref{prop:con}. 
More precisely, this result guarantees that the 2-form $\Omega$ defined by $\widehat{N}=N+\pi^\sharp\Omega^\flat$, 
where $N$ is as in 
\eqref{eq:canten} and $\widehat{N}=\widehat\pi^\sharp{\pi^\sharp}^{-1}$, is closed. Said differently, one can think that the Poisson tensor $\widehat\pi$ described in \cite{NDC-D} can be obtained as a 
deformation of the 
tensor $N$ by a suitable closed 2-form $\Omega$. 
For the reader convenience, we will compute below the 2-form $\Omega$ providing this deformation in the $B_2$ case.

To this end, first we recall \cite[equation (38)]{NDC-D} that
\[
{\widehat\pi}^\sharp=\left(\begin{array}{cc|cc} 0 & 2p_2 & -p_1^2-2{\rm e}^{q_1-q_2} & {\rm e}^{q_1-q_2}-2{\rm e}^{q_2}\\
-2p_2 & 0 & -{\rm e}^{q_1-q_2} & -p_2^2-2{\rm e}^{q_2}\\
\hline
p_1^2+2{\rm e}^{q_1-q_2} & {\rm e}^{q_1-q_2} & 0 & {\rm e}^{q_1-q_2}(p_1+p_2)\\
2{\rm e}^{q_2}-{\rm e}^{q_1-q_2} & p_2^2+2{\rm e}^{q_2} & -{\rm e}^{q_1-q_2}(p_1+p_2) & 0
  \end{array}\right),
\]
which entails that
\[
\widehat{N}=\left(\begin{array}{cc|cc} -p_1^2-2{\rm e}^{q_1-q_2} & {\rm e}^{q_1-q_2}-2{\rm e}^{q_2} & 0 & -2p_2\\
-{\rm e}^{q_1-q_2} & -p_2^2-2{\rm e}^{q_2} & 2p_2 & 0\\
\hline
0 & {\rm e}^{q_1-q_2}(p_1+p_2) & -p_1^2-2{\rm e}^{q_1-q_2} & -{\rm e}^{q_1-q_2}\\
-{\rm e}^{q_1-q_2}(p_1+p_2) & 0 & {\rm e}^{q_1-q_2}-2{\rm e}^{q_2} & -p_2^2-2{\rm e}^{q_2}\end{array}\right).
\]
Hence we obtain
\[
\Omega^\flat=\left(\begin{array}{cc|cc} 0 & -{\rm e}^{q_1-q_2}(p_1+p_2) & p_1^2+2{\rm e}^{q_1-q_2}+p_1 & {\rm e}^{q_1-q_2}\\
{\rm e}^{q_1-q_2}(p_1+p_2) & 0 & 2{\rm e}^{q_2}-{\rm e}^{q_1-q_2} & p_2^2+ 2{\rm e}^{q_2}+p_2\\
\hline
-p_1^2-2{\rm e}^{q_1-q_2}-p_1 & {\rm e}^{q_1-q_2}-2{\rm e}^{q_2} & 0 & -2p_2\\
-{\rm e}^{q_1-q_2} & -p_2^2-2{\rm e}^{q_2}-p_2 & 2p_2 & 0\end{array}\right)
\]
or, equivalently,
\begin{align*}
\Omega=\,&{\rm e}^{q_1-q_2}(p_1+p_2)dq_1\wedge dq_2-(p_1^2+2{\rm e}^{q_1-q_2}+p_1)dq_1\wedge dp_1-{\rm e}^{q_1-q_2}dq_1\wedge dp_2\\
&+({\rm e}^{q_1-q_2}-2{\rm e}^{q_2})dq_2\wedge dp_1-(p_2^2+2{\rm e}^{q_2}+p_2)dq_2\wedge dp_2+2p_2 dp_1\wedge dp_2,
\end{align*}
that a simple computation shows to be closed.
\end{exa}


\end{document}